\documentclass[letterpaper]{article}

\usepackage[T1]{fontenc}

\usepackage{geometry}
\usepackage{setspace}

\usepackage{graphicx}
\usepackage{float}
\usepackage{cite}
\newfloat{scheme}{htbp}{los}
\floatname{scheme}{Scheme}
\floatname{chart}{Chart}
\newfloat{graph}{htbp}{loh}

\usepackage{chemformula} 
\usepackage[version = 4]{mhchem} 
\usepackage{pdfpages}
\usepackage{authblk}
\author[1]{Alice Cartoceti}
\author[2] {Masoumeh Alihosseini}
\author[1]{Gianni Conti}
\author[3,4]{Eliecer Pelaez-Sifonte}
\author[5]{Alessio Orbelli Biroli}
\author[6]{Francesco Sedona}
\author[1]{Paolo D'Agosta}
\author[1]{Francesco Tumino}
\author[1]{Andrea Li Bassi}
\author[3,4]{Jorge Lobo-Checa}
\author[1]{Carlo S. Casari*}
\author[2,7]{Simona Achilli*}
\affil[1]{Department of Energy, Politecnico di Milano, Milano, Italy}
\affil[2]{Department of Physics ‘Aldo Pontremoli', Università degli Studi di Milano, Milano, Italy}
\affil[3]{Instituto de Nanociencia y Materiales de Aragon (INMA), CSIC-Universidad de Zaragoza, Zaragoza, Spain}
\affil[4]{Departamento de Física de la Materia Condensada, Universidad de Zaragoza, Zaragoza, Spain}
\affil[5]{Department of Chemistry, Università di Pavia, Pavia, Italy}
\affil[6]{Dipartimento di Scienze Chimiche, Università degli Studi di Padova, Padova, Italy}
\affil[7]{INFN Sezione di Milano and ‘European Theoretical Spectroscopy Facility' (ETSF), Milano, Italy}

\title{Atomically abrupt sp-sp$^2$ all-carbon heterojunction preserving the electronic identity of constituent allotropes
}
\date{*Email: carlo.casari@polimi.it, simona.achilli@unimi.it}

\begin{document}

\maketitle

\begin{abstract}
Covalently integrating distinct carbon allotropes with atomic precision remains, to date, elusive, despite the clear advantage of combining complementary functionalities while preserving their electronic identities. Here, we realize an all-carbon lateral heterojunction combining ($sp^2$) armchair graphene nanoribbons (GNRs) and $sp$–$sp^2$ hydrogenated graphdiyne (hGDY) on Au(111) through Fe-mediated hierarchical on-surface synthesis. Low-temperature scanning tunnelling microscopy and spectroscopy combined with density-functional theory reveal an atomically abrupt (<0.5 nm), asymmetric type-I interface. Redistribution of an edge-localized GNR state onto the first hGDY unit demonstrates local electronic continuity, whereas their characteristic electronic structures re-emerge immediately beyond the interface. Calculations further indicate that heterojunction formation weakens the interaction of both constituents with the underlying Au(111) substrate, widening their gaps and reducing adsorption-induced interfacial polarization. These results establish lateral covalent integration as a route to engineer atomically defined electronic interfaces and substrate coupling in all-carbon nanostructures.
  
\end{abstract}

\section*{Keywords}
sp--sp$^2$ carbon heterojunctions, graphene nanoribbons, graphdiyne, on-surface synthesis, carbon allotropes, STM, STS, DFT

\newpage



Atomically precise lateral heterojunctions integrating distinct two-dimensional carbon allotropes through covalent interfaces offer a powerful route to engineer the electronic properties of low-dimensional carbon nanostructures. By combining different $\pi$-conjugated building blocks within a single architecture, they can generate interface-specific electronic states and functionalities beyond those of the individual constituents.\cite{serafini2021designing,Xia2015,paupitz2026concise} On-surface synthesis provides the atomic-scale control required to realize such structures.\cite{yin2023surface} To date, however, most carbon heterojunctions have been based on graphene nanoribbons (GNRs), combining segments with different widths, edge topologies, or chemical compositions,\cite{Chen2015Molecular,Rizzo2018Topological,Bronner2018_hierarchical,Zhang2015-Sulfur} while atomically precise heterojunctions between carbon allotropes characterized by fundamentally different hybridization and conjugation schemes remain largely unexplored.

Graphdiyne provides an ideal counterpart to the $sp^2$ network of GNRs, owing to its characteristic combination of $sp$- and $sp^2$-hybridized carbon atoms and diacetylenic linkages. This distinct conjugation gives rise to electronic and structural properties markedly different from those of graphene and has stimulated interest in applications ranging from electronics and optoelectronics to catalysis and molecular sieving.\cite{xue20182d,yu2019graphdiyne} Integrating graphdiyne and GNRs within a covalently bonded architecture could therefore provide access to atomically abrupt interfaces between two fundamentally different carbon $\pi$-networks. However, achieving such all-carbon interfaces is highly challenging. On-surface synthesis of hydrogenated graphdiyne (hGDY) on Au(111) typically results in Au incorporation into the diacetylenic framework, forming organometallic C–Au–C motifs, while residual Br by-products can further hinder the formation of extended hGDY domains and their covalent connection to other carbon nanostructures.\cite{cartoceti2026graphene}

Here, we overcome these limitations and realize atomically precise all-carbon heterojunctions through a hierarchical on-surface synthesis strategy. Building on our finding that Fe promotes both the removal of Br by-products and Au adatoms,\cite{cartoceti2026iron} we extend this approach to covalently integrate two distinct carbon allotropes: armchair graphene nanoribbons (aGNRs) and hGDYs. By combining low-temperature scanning tunneling microscopy and spectroscopy (LT-STM/STS) with density-functional-theory calculations, we resolve the atomic and electronic structure of the resulting $sp^2$/$sp$–$sp^2$ aGNR-hGDY interface. The two lattices form a direct covalent junction without an intermediate phase, while the associated electronic transition occurs within less than 0.5 nm, resulting in an atomically abrupt interface. We further show that the lateral covalent coupling established upon heterojunction formation drives both aGNR and hGDY into a structural and electronic configuration that interacts more weakly with the underlying Au(111) surface than when the two constituents are adsorbed separately. This weaker substrate coupling partially restores the electronic features characteristic of the corresponding freestanding systems, suggesting a strategy in which lateral covalent coupling is used to control vertical interaction with the supporting substrate.

\section{Fe-mediated synthesis of aGNR--hGDY junctions}
\begin{figure}[h!]
\centering
\includegraphics[width=0.9\textwidth]{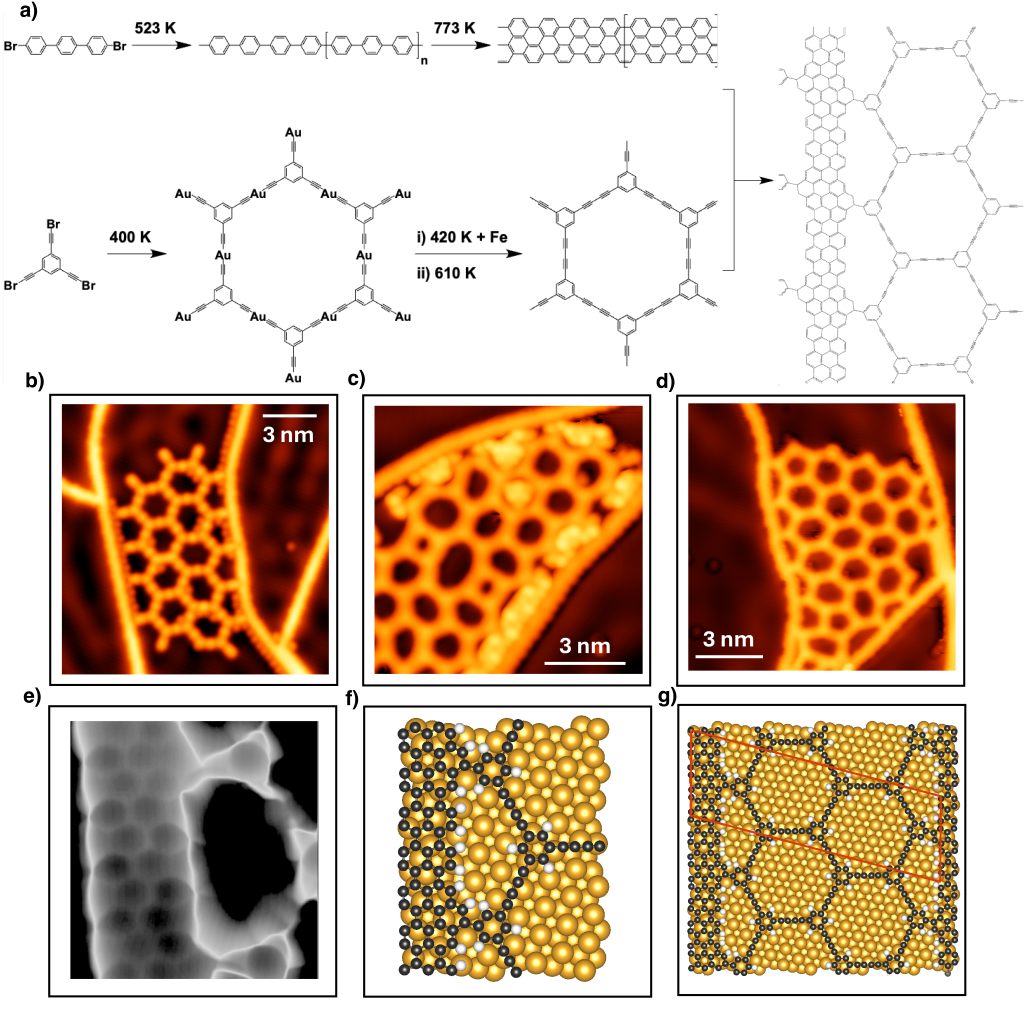}
\caption{a) Hierarchical synthetic route model to achieve aGNR-hGDY heterostructures. LT-STM images of b) the aGNR-hGDY system on Au(111) upon annealing at 400 K, before the aGNR-hGDY bonding \cite{cartoceti2026graphene}; c) the system upon Fe deposition and annealing at 450 K, with the formation of FeBr$_2$ aggregates and removal of Au adatoms from the hGDY; d) exemplified covalent all-carbon heterojunctions between the hGDY and 3-aGNRs, obtained upon annealing at 610 K; e) an exemplifying close-up of the junction covalent bond between hGDY and a 6-aGNR, obtained with CO-functionalized tip. Grey color scale has been chosen to better underline the structural features at the junction. f) model of the heterostructure bonding. g) structural model of the heterojunction and unit cell (red line). STM set-point: (b) -10 mV/80 pA, (c) -10 mV/8 pA, (d) -20 mV/40 pA, (e) -3mV/60pA.
}\label{Fig1}
\end{figure}

Covalent aGNR--hGDY heterostructures (HTSs) were synthesized on Au(111) through a sequential, multi-precursor on-surface process mediated by Fe, as summarized in Figure~\ref{Fig1}a (details in Figure~S1 and Methods). Low-coverage aGNRs, predominantly 3- and 6-aGNRs, were first prepared from 4,4''-dibromo-\textit{p}-terphenyl (DBTP) following an established procedure \cite{basagni2015molecules,de2020templating,merino2017width}. The nanoribbons subsequently served as templates for the growth of metalated hGDY islands from 1,3,5-tri(bromoethynyl)benzene (tBEB). At this stage, residual Br species accumulate predominantly at the aGNR--hGDY boundary and form complexes with Au adatoms at the hGDY terminations, preventing direct covalent coupling (Figure~\ref{Fig1}b) \cite{cartoceti2026graphene}. Fe deposition followed by annealing at 450 K promotes the formation of FeBr$_2$ species and the removal of Au adatoms from the network, leading to complete hGDY demetallation (Figure~\ref{Fig1}c) \cite{cartoceti2026iron}. Subsequent annealing at 530 K and 610 K induces covalent coupling to the aGNRs, the removal of FeBr$_2$, and partial reordering of the hGDY network (Figure~\ref{Fig1}d) \cite{cartoceti2026iron}. This Fe-mediated process increases the linking efficiency to $97.47\pm0.88$\%, compared with $\sim41$\% for thermally driven synthesis and $\sim66$\% following atomic-hydrogen exposure \cite{cartoceti2026graphene}. The corresponding statistical analysis is reported in Supplementary Section S2.

Line-profile 
analysis performed over several LT-STM images of the hGDY upon the formation of the all-carbon heterojunction yields an average hGDY periodicity of $1.58\pm0.05$~nm
At the junction, the two carbon atoms of a terminal hGDY acetylene unit bond to two edge carbon atoms of the aGNR, forming an additional aromatic ring (Figure~\ref{Fig1}e,f) \cite{cartoceti2026graphene} with a preferred ($\sim85.11\pm2.02$\%) perpendicular linking between the two subsystems (see S2 for statistical details).


Guided by these observations, we built the atomistic model, shown in Figure~\ref{Fig1}g, reproducing the experimentally observed aGNR--hGDY geometry while maintaining commensurability with Au(111); its complete geometrical description is provided in Methods.

\section{Atomically abrupt type-I interface}

To determine how the electronic states of aGNR and hGDY evolve across their covalent boundary, we combined spatially resolved LT-STS measurements with density-functional-theory calculations.

\begin{figure}[h!]
\centering
\includegraphics[width=0.85\textwidth]{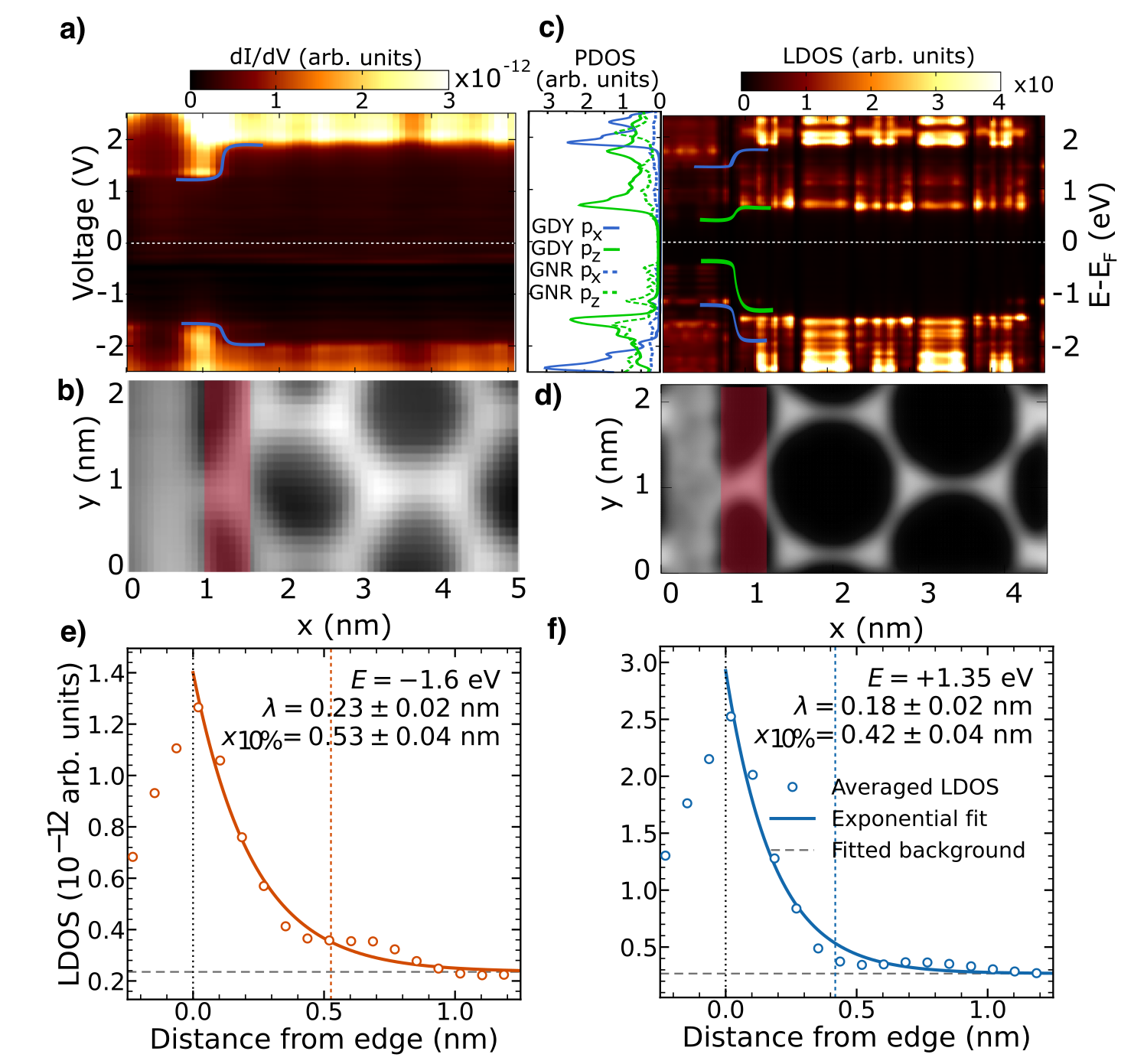}
\caption{a) Experimental averaged LDOS across the heterojunction. Blue lines mark the experimental band edges. b) Atomically resolved LT-STM image of the area used to average the signal reported in a). Shaded red area marks the interface extension. (c) Calculated LDOS across the heterojunction averaged over a unit cell. Inset: projected density of states (PDOS) on $p_z$ and $p_{x,y}$ states on carbon. Blue lines mark the band edges corresponding to $p_{x,y}$ states. Green lines evidence the true band edges across the junction due to $p_z$ states. (d) Constant-current simulated STM image of the heterojunction. Shaded red area marks the interface extension. e), f) Experimental fit of the dI/dV profiles extracted at -1.6 eV (interface state), and 1.35 eV (unoccupied band), respectively. Horizontal dashed lines indicate the fitted background, while vertical colored dashed lines mark the distance at which the spectral weight above background decreases to 10\% of its value at the junction. STS set-point: -0.55 V/60 pA, with lock-in amplitude 20 mV and frequency 817 Hz.
}\label{Fig2}
\end{figure}

Figure \ref{Fig2}a shows the experimental local density of states (LDOS) along the heterojunction, obtained by averaging the dI/dV signal, acquired in constant-current mode, along the 2 nm vertical region parallel to the aGNR axis shown in the STM image in Figure \ref{Fig2}b, with a step of 0.2 nm. The corresponding calculated LDOS along the junction, averaged over the same spatial interval (simulated STM image in Figure \ref{Fig2}d) is reported in Figure \ref{Fig2}c.
The experimental profile displays prominent spectroscopic features at approximately -1.6 V and +1.35 V in the aGNR region and at approximately -2.0 V and +2.1 V in the hGDY region. Their relative alignment, highlighted by blue lines, is consistent with a type-I heterojunction, in which the experimentally accessible gap of the aGNR lies within that of hGDY. A comparable band offset is reproduced by the calculated LDOS, which exhibits intense features localized on the aGNR and hGDY domains at energies closely matching the experimental values (blue lines in Figure \ref{Fig2}c), corresponding to $p_{x,y}$ states (left inset of Figure \ref{Fig2}c).

Calculated electronic structure, however, reveals additional states closer to the Fermi level. The calculated frontier orbitals delimiting the aGNR bandgap are found approximately at -0.5 eV and +0.3 eV, whereas those delimiting the hGDY bandgap are predicted at approximately -1.55 eV and +0.5 eV, as indicated by the green markers in Figure \ref{Fig2}c. The resulting alignment therefore retains a type-I character, but is markedly asymmetric, with a large valence-band offset and nearly aligned conduction-band edges. This electronic landscape strongly confines valence states to the aGNR, while providing a substantially lower barrier for conduction-band states across the junction. The proximity of the hGDY conduction-band edge to the common Fermi level further indicates a pronounced n-type character of the hGDY domain, while aGNR shows a mild p-type character.
The PDOS on carbon orbitals (inset of Figure \ref{Fig2}c) reveals that the states resolved by the calculation near the Fermi level have $p_z$ character. These states are not clearly resolved in the spatially averaged experimental profile
mainly due to the difference in the relative intensity between the $p_{x,y}$ and $p_{z}$ states determined by the setpoint chosen for the STS measurements \cite{cartoceti2026graphene}.
However, the experimental validation of their presence is given by local dI/dV spectra and energy-resolved conductance maps discussed below.
Both the experimental and calculated LDOS (Figure \ref{Fig2}a, c, respectively) also reveal a sharp state at -1.6 eV localized at the interface between the two subsystems, which rapidly decays in the hGDY.

An exponential fit of the spatially resolved intensity of dI/dV profiles (Figure \ref{Fig2}e-f) yields comparable decay lengths of $\sim0.2$ nm for the experimentally observed aGNR band features, with their spectral weight decreasing to 10\% of the interfacial value within approximately 0.5 nm (red shaded area in Figure \ref{Fig2}b,d). This subnanometer spatial evolution, leading to a rapid recovery of the characteristic electronic fingerprints of each allotrope away from the boundary, demonstrates the electronically abrupt nature of the heterojunction and excludes an extended electronic transition region.
\begin{figure}[h!]
\centering
\includegraphics[width=0.85\textwidth]{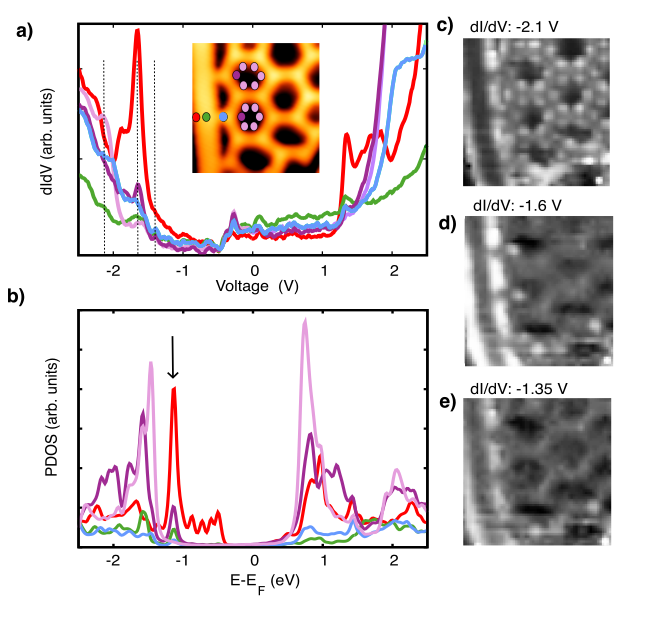}
\caption{a) Conductance spectra acquired at the colour-coded dot positions reported in the LT-STM image in the inset. b) Projected density of states on carbon atoms in the heterostructure corresponding to the positions marked in panel a). c-e) Energy-dependent dI/dV maps at different voltages reported above each panel. STS set-point: -0.55 V/60 pA, with lock-in amplitude 20 mV and frequency 817 Hz.
}\label{Fig3}
\end{figure}

Site-resolved experimental STS spectra (Figure~\ref{Fig3}a), compared with the corresponding calculated PDOS (Figure~\ref{Fig3}b and Figure S3) and supported by experimental dI/dV maps (Figure~\ref{Fig3}c-e and Figure S4-S5), reveal the spatial distribution and orbital character of the relevant states in the heterojunction.
STS spectra measured in different positions of the heterostructure (Figure \ref{Fig3}a, color code in the inset) show a prominent feature at -2.1 V that extends over the edges of aGNR (red line), the bulk hGDY (violet line) and with a lower weight also at the interface (cyan and purple lines) and in the inner aGNR (green line), in fair agreement with the calculated PDOS (features at $\sim$ -2.1 eV in Figure \ref{Fig3}b). The experimental dI/dV map at that energy (Figure \ref{Fig3}c) allows us to identify this state as a $p_z$ state at the edge of the aGNR (lobes protruding outwards) and $p_{x,y}$ state on the hGDY region (lobes in plane), as confirmed by the orbital-projected PDOS analysis (Figure S3 b,f,g).
Another intense feature at -1.6 V in the STS spectrum is mainly localized at the aGNR edge and on the inner carbon atom of the first phenyl ring of the hGDY (red and purple lines), in agreement with the calculated PDOS (red and purple lines in correspondence of the arrow at -1.15 eV). As further confirmed by the experimental dI/dV map in Figure \ref{Fig3}d, this state at the edge of aGNR is interrupted at the linking site of the junction but, rather than disappearing at the ribbon termination, the edge-derived wavefunction extends into the first hydrogenated phenyl unit of the hGDY domain, where its intensity is slightly reduced.
This spatial continuity of the interface state is consistent with covalent coupling across the junction.


Experimental dI/dV maps acquired in the energy range between -1.7 V and -0.7 V (Figure S4), with the map in Figure \ref{Fig3}e as a representative example at -1.35 V, reveal the presence of $p_z$-derived states on the hGDY network, providing experimental support for the theoretical prediction in Figure \ref{Fig2} that $p_z$-derived states also contribute to its electronic structure. 
In the unoccupied part of the spectrum, states with the same characteristics as those at -1.6 V and -2.1 V are found respectively at 1.35 V and 2.1 V in the experiment, representing the symmetric counterpart of the interface state and $p_{x,y}$ manifold (Figure S5), as confirmed by the calculations (states at 1.0 eV and 2.0 eV in Figure \ref{Fig3}b). As already noted, the theoretical PDOS also displays deeper unoccupied hGDY p$_z$ states near the Fermi level.

\section{Effect of heterostructure on constituent–substrate coupling}
Having characterized the electronic states associated with the covalent junction, we next examine whether the heterostructure formation modifies the interaction of aGNR and hGDY with Au(111).

The covalent HTS weakly perturbs the substrate, as indicated by the complete recovery of the Au(111) herringbone reconstruction following the hGDY conversion from the metalated to the covalent phase (Figure~S6a,b), also observed for isolated covalent hGDY \cite{cartoceti2026iron}. Consistently, the Au(111) surface state onset beneath the HTS lies at $-0.38$~eV, only 0.07~eV above its position on pristine Au(111) ($-0.45$~eV) \cite{andreev2004adsorbed}. This shift is smaller than the 0.10~eV measured beneath isolated covalent hGDY \cite{cartoceti2026iron}, suggesting that heterostructure formation further attenuates substrate coupling. Substantially larger shifts occur beneath strongly interacting metalated $sp$--$sp^2$ networks (Figure~S6c) \cite{li2024scanning,d2025unraveling,yang2020metalated}.

\begin{figure}[h!]
\centering
\includegraphics[width=1.0\textwidth]{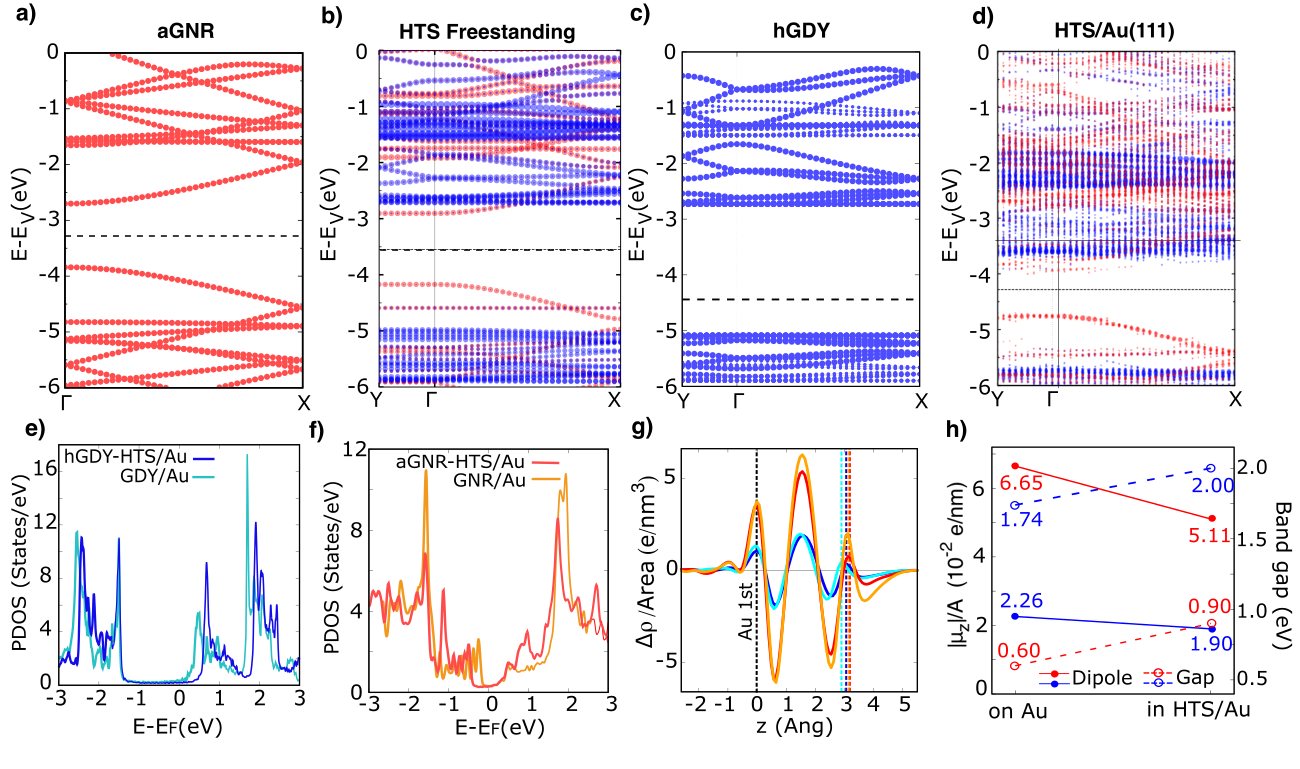}
\caption{a) Band structure of the freestanding aGNR in a \(4\times1\) supercell to match its periodicity in the HTS. b) Band structure of the freestanding aGNR-hGDY heterostructure.  c) Band structure of the freestanding hGDY in the same oblique cell as in the HTS. d) Band structure of the HTS on Au(111). In red and blue the aGNR and hGDY contribution, respectively. e) Comparison between the PDOS of pristine hGDY/Au(111) and the PDOS of a hGDY unit cell in the central region of the HTS. f) Comparison between the PDOS of pristine aGNR/Au(111) and of an aGNR unit in HTS. g) Surface-averaged induced charge along the direction perpendicular to the HTS. h) Change of the local dipole density compared to the change in the energy gap. Red and blue refer to hGDY and aGNR, respectively.
}\label{Fig4}
\end{figure}

To disentangle the effects of lateral covalent coupling and substrate adsorption, we compare the electronic structures of the isolated constituents, the freestanding HTS and the corresponding Au-supported systems. The band structures of freestanding aGNR and hGDY (Figure~\ref{Fig4}a,c) reveal their distinct work functions and band gaps, providing a reference for identifying the changes induced by heterojunction formation and interaction with Au(111).

Upon covalent coupling into the freestanding aGNR-hGDY heterostructure (Figure \ref{Fig4}b), the common Fermi level shifts according to the new work function of the combined system, while the electronic manifolds of the two constituents remain largely preserved with a larger shift of the hGDY bands with respect to the common Fermi level, compared to aGNR. The band projection onto the aGNR and hGDY subunits (red and blue in Figure \ref{Fig4}b, respectively), shows that the vast majority of the bands retain their original character. The only exception is the nearly flat interface band at -4.55 eV in Figure \ref{Fig4}b (-1.15 eV from the Fermi level), which displays mixed aGNR–hGDY character, consistent with its localization at the covalent junction. The effect of the Au(111) substrate can be understood by considering the behaviour of the individual constituents on the surface. Isolated aGNR on Au(111) exhibits a predominantly p-type band alignment \cite{merino2017width}, whereas hGDY displays the opposite n-type character \cite{cartoceti2026iron}. Upon adsorption of the heterostructure (Figure~\ref{Fig4}d) the bands shift to align with the Au(111) work function. The two constituent-derived manifolds remain clearly distinguishable, and qualitatively preserve their dispersion and the different Fermi-level alignment, indicating that the weak interaction with the substrate does not suppress the electronic identity of the individual subunits.

A more quantitative insight is obtained from the comparison between carbon-projected density of states calculated on hGDY and aGNR units in the HTS, 1 nm away from the covalent junction (Figure~\ref{Fig4}e,f, blue and red lines, respectively), and the corresponding isolated systems adsorbed on Au(111) (Figure~\ref{Fig4}e,f, cyan and orange lines, respectively). Remarkably, both the constituents exhibit a wider projected band gap within the heterostructure than when individually supported on Au. The recovery occurs through different band-edge shifts: in hGDY, the gap widens predominantly through an upward shift of the conduction-band edge, whereas in the aGNR it originates mainly from a downward displacement of the valence-band edge. Since these band edges are those most strongly perturbed by the Au substrate in the isolated systems, the observed behaviour suggests that the formation of the lateral heterostructure modifies the interaction of both constituents with Au(111), partially restoring a more freestanding-like electronic structure.

The gap increase can be related to the larger adsorption height of the carbon network above Au. In the heterostructure, hGDY lies approximately 0.2~\AA~farther from the substrate than isolated hGDY, whereas the change in the GNR height is smaller (0.1~\AA). However, because the isolated GNR intrinsically prefers a larger adsorption distance than hGDY, the covalent junction forces the hGDY network to follow the higher equilibrium position imposed by the GNR. The slight additional lifting of the GNR may be related to the electronic reconstruction at the junction, where interface states mix with the GNR edge states and modify its interaction with Au.

Likewise, the amount of adsorption-induced integrated charge in the constituent regions away from the junction decreases moderately upon HTS formation (approximately 12\% for hGDY and 10\% for the aGNR). Further insight into the modified interaction with the substrate is provided by the spatially resolved adsorption-induced charge-density difference \(\overline{\Delta\rho}_{R}(z)\) and local dipole density (see Methods).
Figure~\ref{Fig4}g reports \(\overline{\Delta\rho}_{R}(z)\) for the hGDY and aGNR regions when individually adsorbed on Au(111) (cyan and orange, respectively) and when incorporated into the heterostructure (blue and red, respectively), considering in the latter case regions sufficiently far from the covalent junction. Besides the substantially larger charge redistribution observed for the aGNR, the profiles show small shifts in the accumulation and depletion regions, particularly in the features associated with the carbon layer, consistently with the small changes in adsorption height upon HTS formation (position marked by vertical coloured dashed lines).

Upon incorporation into the HTS, the local dipole density decreases from $2.26\times10^{-3}$ to $1.90\times10^{-3}$~e/\AA\ for hGDY and from $6.65\times10^{-3}$ to $5.11\times10^{-3}$~e/\AA\ for aGNR, corresponding to reductions of 16\% and 23\%, respectively (Figure~\ref{Fig4}h). In both constituents, this attenuation accompanies partial gap recovery. 
The larger dipole reduction of the aGNR, despite its smaller vertical displacement, shows that interfacial polarization is not governed solely by adsorption height. Instead, it is consistent with the junction-induced redistribution of GNR-derived $\pi$-state spectral weight across the covalent link and onto the first hGDY benzene ring, possibly reflecting an extension of the GNR aromaticity. The associated change in the overlap of edge-derived states with Au provides a plausible electronic contribution to the reduced GNR–Au polarization.
For hGDY, the larger upward displacement provides a more direct structural mechanism for weaker Au coupling and gap widening.

\section{Conclusions}
In conclusion, we have demonstrated the atomically precise on-surface synthesis of the atomically precise all-carbon lateral heterojunction integrating carbon allotropes with fundamentally different bonding networks, namely an sp$^2$ graphene nanoribbon and an sp–sp$^2$ hydrogenated graphdiyne. The Fe-assisted synthetic strategy enables the formation of a fully covalent interface while preserving the structural order of both constituents.
Low-temperature STM/STS combined with first-principles calculations reveals that the heterojunction is electronically abrupt at the atomic scale, with the interface extending less than 0.5 nm at the covalent link. The band alignment exhibits an asymmetric type-I character that may provide a route toward carrier-selective confinement in all-carbon heterojunctions.
At the same time, the electronic identity of the two carbon allotropes remains well preserved away from the interface, with the direct hybridization restricted to the immediate vicinity of the junction.
Remarkably, despite the highly localized nature of the electronic coupling, the formation of the heterostructure modifies the interaction of both constituents with the underlying Au(111) substrate. This results in a constituent-dependent reduction of the substrate-induced electronic renormalization and interfacial polarization, leading to a partial recovery of their intrinsic electronic structures while maintaining direct contact with the metallic support.
The decoupling induced by the heterojunction formation should be regarded as an additional effect superimposed on the stronger decoupling already produced by the organometallic-to-covalent transformation of hGDY. Although smaller in magnitude, this additional reduction in substrate coupling is conceptually relevant because it affects both constituents and extends beyond the immediate junction region.
Beyond the specific aGNR–hGDY system investigated here, this heterostructure provides a proof of concept for the atomically precise integration of other carbon allotropes with distinct bonding motifs and electronic properties using analogous on-surface synthetic strategies. More broadly, these findings establish lateral covalent heterostructuring as an effective approach not only for engineering electronic states at atomically precise carbon interfaces, but also for tailoring the interaction of distinct carbon allotropes with metallic substrates, opening new opportunities for the design of functional all-carbon quantum nanostructures.
\newpage
\section{Methods}\label{sec11}

\subsection*{Experiment}
All the experiments were carried out under ultra-high vacuum conditions, with a base pressure below $10^{-10}$~mbar. Single-crystal Au(111) (MaTeck GmbH) was employed for the on-surface synthesis, following several cleaning cycles of Ar$^+$ sputtering and annealing at 770~K. Molecular precursors, i.e. 1,3,5-tri(bromoethynyl)benzene (tBEB), and 4,4''-dibromo-p-terphenyl (DBTP), were loaded in powder form in two separate Knudsen cells and were deposited on Au(111) by thermal evaporation, after calibration of the molecular flux with a quartz microbalance.
The lateral heterostructure was obtained through the sequential deposition of DBTP and tBEB molecules on the Au(111) substrate, held at room temperature in front of the evaporator. To first obtain aGNRs, DBTP molecules were sublimated on Au(111) for 16 minutes, with a rate of about 0.025 ML/minute, keeping the crucible temperature set to 353~K, reaching a chamber pressure of 1.6~$\times~10^{-8}$~mbar. Upon evaporation at RT, DBTP molecules form an ordered self-assembly of unreacted monomers stabilized by lateral hydrogen bonds between Br and H terminals \cite{basagni2015molecules}. It is known that the debromination of DBTP molecules occurs in the 360-400 K range, and Br desorption from Au(111) surface at about 650 K \cite{basagni2015molecules}. Upon annealing at 523 K for 30 minutes, poly(p-phenylene) (PPP) chains are formed \cite{basagni2015molecules}. Further annealing of the sample at 773 K for 30 minutes promotes the cyclodehydrogenation of PPP chains, and their coupling into aGNRs with polydisperse widths, from 3- to 6- and 9-aGNR, and interconnecting branches \cite{merino2017width,cartoceti2026graphene}. tBEB molecules are evaporated for about 25 seconds over the obtained aGNR/Au(111) sample, keeping the crucible at 300 K, reaching a chamber pressure of 2~$\times~10^{-8}$~mbar. Upon surface-catalyzed halogen cleavage, tBEB molecules undergo organometallic coupling reaction through the incorporation of gold adatoms from the substrate, generating metalated hGDY in between the nanoribbons \cite{sun2016dehalogenative,d2025unraveling, yang2020metalated, shu2020atomic, rabia2020structural,cartoceti2026graphene}. To improve the order of the hGDY, a mild annealing at about 400 K is performed. 
At this point, Fe atoms are evaporated on the obtained system through an e-beam evaporator equipped with an integral flux monitor. The sample is maintained at room temperature while Fe atoms are deposited on top, setting the emission current I$_{em}$ to 13.7 mA, the ion flux to 0.1 nA, and the deposition time to 80 seconds. A constant pressure of 2.5$\times10^{-10}$~mbar during Fe evaporation ensures the cleanliness of the deposit, without contamination from the metallic rod. After Fe deposition, a progressive annealing of the system from 370 K to 610 K, with an increase in the set current of 0.2 A upon each 20-minute step, promotes the bonding between the covalent hGDY and the aGNRs.
Low-temperature scanning tunneling microscopy and spectroscopy (LT-STM/STS) investigations were conducted at the Laboratorio de Microscopías Avanzadas of the Universidad de Zaragoza, with a Scienta Omicron microscope, cooled down to 4.8 K, using a W tip. LT-STS spectra and maps were obtained through a lock-in amplifier with an oscillation frequency of f$_{osc}$ = 817.3 Hz and a modulation amplitude of V$_{RMS}$ = 20 mV.
STM images were analysed using Gwyddion 2.63 \cite{nevcas2012gwyddion}. The hGDY periodicity was determined from line profiles acquired over several LT-STM images and from the corresponding two-dimensional fast Fourier transforms. Junction yields and relative populations of the $90^\circ$ and $60^\circ$ configurations were obtained by statistical analysis of multiple STM images, as detailed in Supplementary Section S2. STS data were analysed using Igor Pro and SpectraFox.


\subsection*{Theory}
Density functional theory (DFT) calculations were performed using the SIESTA code \cite{Sole02} that features a pseudopotential description of the electron–ion interaction and a localized atomic-orbital basis set. A mesh cutoff of 400 Ry was adopted, and exchange-correlation effects were treated within the GGA using the PBE functional \cite{PBE}. The interaction between the covalent network and the Au substrate was described by including van der Waals corrections through the Grimme dispersion scheme (Grimme D3) \cite{Grimme2010D3}.

The structural model of the heterostructure consists of one hGDY repeated unit (theoretical periodicity of 1.65~nm) matched to four aGNR unit cells (theoretical periodicity of 0.43~nm), resulting in an intrinsic lattice mismatch of approximately 5\%. To achieve simultaneous epitaxial matching with the underlying Au(111) substrate, the common periodic direction of the hGDY and the aGNR was rotated by $8.81^\circ$ with respect to the Au(111) $[1\bar{1}0]$ direction and slightly rescaled to 1.66~nm. This introduces only 0.8\% tensile strain in the hGDY and 4\% compressive strain in the aGNR. The second lattice vector, running across the heterojunction, is oriented $5.12^\circ$ away from the Au(111) [$11\bar{2}$] direction and has a length of 49.21~\AA{}, accommodating one 6-aGNR and two complete hexagonal units of the hGDY network. The resulting oblique supercell, characterized by an internal angle of $\gamma=75.84^\circ$, is fully commensurate with the Au(111) substrate and is described by the epitaxial matrix

\[
\mathbf{M}=
\begin{pmatrix}
5 & 1\\
-7 & 18
\end{pmatrix}.
\]

Structural relaxation preserves the covalent architecture of the $sp$--$sp^2$ heterojunction.

Brillouin-zone sampling was performed using a $3\times2\times1$ k-point mesh for the self-consistent calculations, while a five-times denser mesh was employed for the projected density of states (PDOS). The PDOS is calculated with a broadening of 0.01 eV. 

Simulated STM images were generated within the Tersoff-Hamann approximation \cite{Tersoff}, in both constant-height and constant-current modes, by integrating the electronic states over the energy interval [-0.5,E$_F$]. 

The adsorption-induced charge-density difference was evaluated as
\begin{equation}
\Delta \rho(\mathbf r)=\rho_{\mathrm{C/Au}}(\mathbf r)
-\rho_{\mathrm{C}}(\mathbf r)
-\rho_{\mathrm{Au}}(\mathbf r),
\end{equation}
where $\rho_{\mathrm{C/Au}}$ is the charge density of the adsorbed carbon system (hGDY or aGNR) on Au, while $\rho_{\mathrm{C}}$ and $\rho_{\mathrm{Au}}$ are the charge densities of the isolated carbon system and Au(111) substrate, respectively, evaluated in the same atomic geometries as in the adsorbed case.

We then calculated the laterally averaged ($\overline{\Delta \rho}_{R}$) 
\begin{equation}
\overline{\Delta \rho}_{R}(z) = \frac{1}{A_R} \int_{R} \Delta \rho(\mathbf r) dx dy
\end{equation}
where \(R\) denotes the selected constituent region and \(A_R\) its lateral
area of the unit cell. 
The  effective local adsorption-induced dipole density was
calculated as
\begin{equation}
p_R
=
\frac{\mu_{z,R}}{A_R}
=
\int
\left(z-z_{\mathrm{Au}}\right)
\overline{\Delta\rho}_{R}(z)\,dz,
\label{eq:local_dipole}
\end{equation}
where \(z_{\mathrm{Au}}\) is the position of the topmost Au layer. The regions
were selected sufficiently far from the covalent linking site to minimize the
contribution of direct aGNR-hGDY hybridization.
\bibliographystyle{unsrt}
\bibliography{biblio}

\section{Supporting Information available}
Supporting information reports additional details abouth the synthesis process and the characterization of the electronic properties of the heterojunction.

\section{Acknowledgments}
A.C., A.L.B. and C.S.C acknowledge funding by: Funder: project funded under the National Recovery and Resilience Plan (NRRP), Mission 4 Component 2 Investment 1.3 - Call for tender No. 341 of 15.03.2022 of Ministero dell'Universit\'a e della Ricerca (MUR); funded by the European Union NextGenerationEU - Award Number: project code PE0000021, Concession Decree No. 1561 of 11.10.2022 adopted by Ministero dell’Universit\'a e della Ricerca (MUR), CUP D43C22003090001, Project title ``Network 4 Energy Sustainable Transition NEST' '. F.S. acknowledges the financial support from the University of Padova through the grant  P-DiSC\#02BIRD2024-UNIPD  VS-PpyTM. J.L.C. and E.P.S. acknowledge financial support from the Spanish Ministry of Science and Innovation (Grant PID2022\_138750NB-C21 and ``Severo Ochoa' ' Programme for Centres of Excellence in R\&D CEX2023-001286-S funded by MCIN/AEI/ 10.13039/501100011033 and by ERDF ``A way of making Europe' ' ), from the regional Governments of Aragon (E12\_23R) and from the European Commission (project ULTIMATE-I, grant ID 101007825). J.L.C. and E.P.S. further acknowledge the use of Servicio General de Apoyo a la Investigaci\'on-SAI and the Laboratorio de Microscopías Avanzadas of the Universidad de Zaragoza. S.A. acknowledges the CINECA award under the ISCRA initiative, for the availability of high-performance computing resources and support, and the financial support from INFN (Progetto iniziativa specifica ``TIME2QUEST''). A.O.B. acknowledges support from the Ministero dell’Universit\'a e della Ricerca (MUR) and the University of Pavia through the program ``Dipartimenti di Eccellenza'' 2023--2027. Mario C.G. Passeggi (Jr.) is acknowledged for contributing to the experiment and for the useful discussion.

\section{Author contributions}
A.C. conceptualized and conducted the STM/STS experiments, analyzed the corresponding data, and wrote the initial draft. M. A. performed the DFT calculations. A.C., E.P.S. and J.L.C. conducted LT-STM/STS experiments. A.O.B. synthesized the tBEB molecular precursor. G.C., P.D., F.T. contributed to preliminary on-surface synthesis experiments. J.L.C., F.S., A.L.B., C.S.C. contributed to the supervision and discussion of results. C.S.C. conceptualized the work. S.A. conceptualized and coordinated the theoretical modeling and related analysis and wrote the initial draft. All authors contributed to the revision and final discussion of the manuscript.

\section{Competing interests}
The authors declare no competing interests.


\clearpage
\clearpage
\includepdf[pages=-]{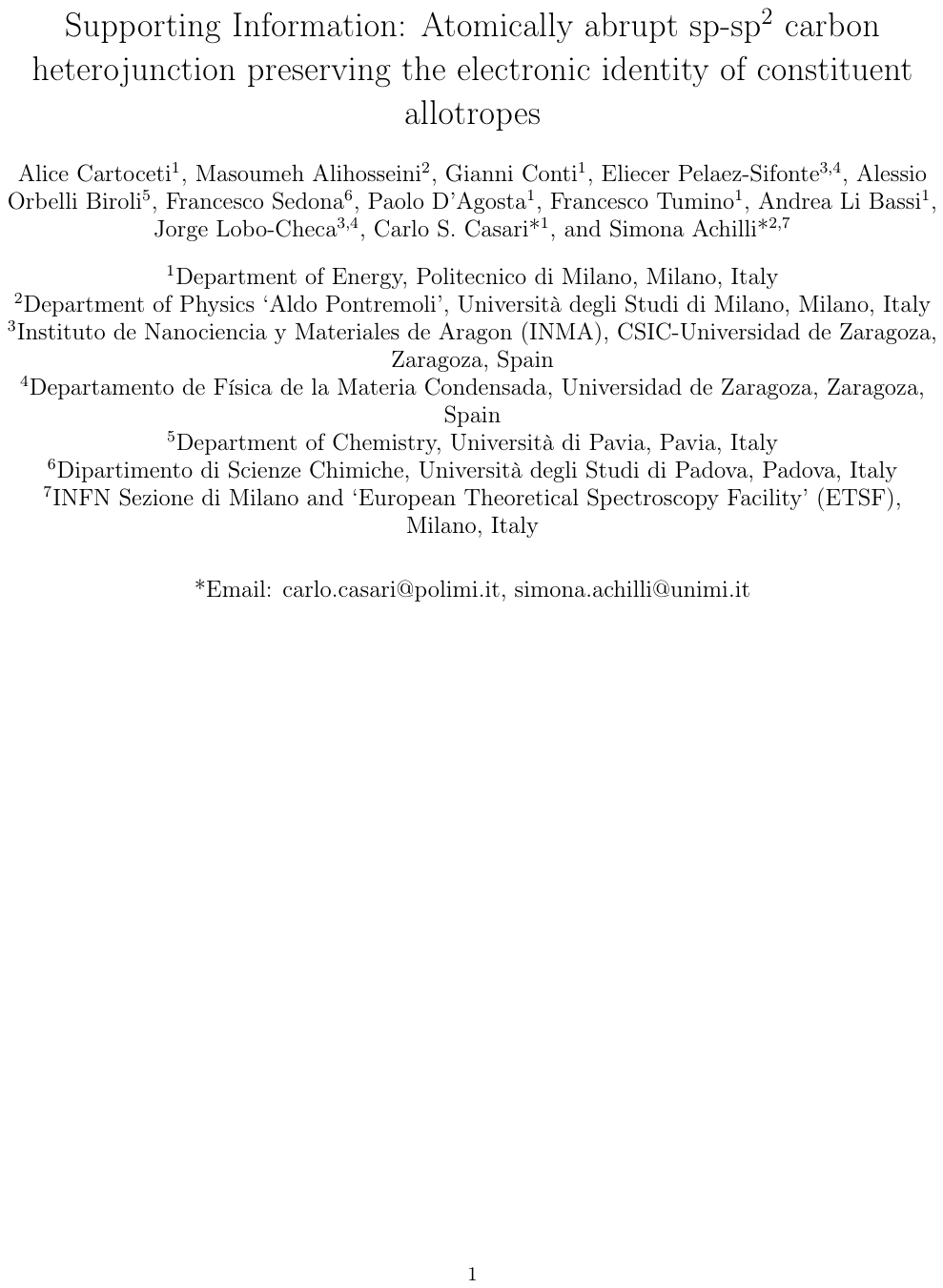}

\end{document}